# Olive oil's polyphenolic metabolites: from their influence on human health to their chemical synthesis

Filipe Monteiro-Silva

[a]Departamento de Química e Bioquímica, Faculdade de Ciências, Universidade do Porto, Rua do Campo Alegre, 4169-007 Porto, Portugal

____________________________________________________________________

English version of abstract

## *Abstract*

A growing number of scientific researches have been demonstrating that olive oil operates a crucial role on the prevention of cardiovascular and tumoral diseases, being related with low mortality and morbidity in populations that tend to follow a Mediterranean diet. Amongst its minor components, polyphenols have been subject to several clinical trials that have established health benefits associated to their antioxidant, antitumoral and anti-atherosclerotic activity.

However the biological activity of polyphenols is dependent not only of their absorption but also of their metabolization. Bioavailability studies have demonstrated that after olive oil intake, not only hydroxytyrosol but also its metabolites can be found in plasma, such as 3,4-dihydroxyphenylacetaldehyde, 3,4-dihydroxyphenylacetic acid, homovanillyl alcohol and its glucoronides.

-----

Portuguese version of abstract

## *Resumo*

Um número crescente de pesquisas científicas demonstram que o azeite opera um papel crucial na prevenção de doenças cardiovasculares e doenças tumorais, estando relacionado com a baixa mortalidade e morbidade em populações que tradicionalmente seguem uma dieta Mediterrânica. De entre os seus componentes minoritários, os polifenóis têm vindo a ser alvo de estudos clínicos que demonstraram o seu benefício para a saúde pela atividade antioxidante, anti-tumoral e anti-aterosclerótica que possuem.

No entanto a atividade biológica destes polifenóis é dependente não só da sua absorção como poderá também depender da sua metabolização. Estudos de biodisponibilidade demonstraram que após o consumo de azeite encontram-se no plasma não só hidroxitirosol, mas igualmente os seus metabolitos, tais como o ácido 3,4-dihidroxifenilacético, o 3,4-dihidroxifenilacetaldeído, o álcool homovanílico e os respectivos glucoronídeos.

## Índice

## 1. Introdução e enquadramento geral

A utilização do azeite para diversos fins é milenar e remonta à Idade de Bronze (3000-600 a.C.) e entre as suas principais aplicações distingue-se obviamente, a utilização na alimentação, mas a sua polivalência permitiu ser usado quer como combustível, quer como unguento para diversas maleitas. O valor deste óleo era tão grande que chegou mesmo a ser usado como prémio nas conquistas efetuadas nos Jogos Olímpicos antigos (Gigon *et al.*, 2010; Boskou, 2006).

Apesar de atualmente ser utilizado na produção de cosméticos e desempenhar um papel importante em diversas religiões, o uso como alimento prevaleceu sobre os restantes nomeadamente como tempero em saladas ou em pratos principais, ou como parte integrante na preparação de sopas, características da dieta mediterrânica.

### 1.1. Oliveiras, azeitonas e azeite

A oliveira, de designação científica *Olea europaea L.*, é uma planta de climas e temperaturas amenas, é resistente à seca e a condições adversas, apesar de as suas raízes crescerem tendencialmente na horizontal e não na vertical. As suas folhas são estreitas, ovalizadas e persistentes – em períodos de cerca de 3 anos – com pequenos pêlos que protegem as mesmas contra a desidratação.

Após o aparecimento das flores, pequenas, de cor amarela/branca, em pequenos aglomerados, ocorre a formação dos primeiros frutos, os quais amadurecem em períodos longos e lentos. As fases de amadurecimento são diversas, dependendo também da variedade, que num primeiro estado de maturação adquirem a cor verde e o seu tamanho final. Posteriormente, a coloração verde dá lugar às fases designadas por "manchada", "púrpura" e "preta". É durante estas transições que a clorofila é gradualmente substituída por antocianinas (Boskou, 2006).

O azeite, por seu lado, é extraído das azeitonas, e as suas qualidades dependem do estado de maturação do fruto, da forma de extração, mas também do tipo de clima, tratamento e cuidados com as oliveiras. Mesmo sem qualquer tipo de cuidado específico, esta planta é capaz de produzir azeitonas, resistindo a climas agrestes, chegando a haver casos reportados de árvores com centenas e milhares de anos.

A sua resiliência permitiu-lhe adaptar-se à bacia Mediterrânica – e apresentar resistência ao sal - e tornar-se numa das plantas de fruto mais produtivas desta área.

*1.2. O azeite e a economia*

No quadriénio 2005–2008, a nível mundial, a produção de azeitonas foi cerca de 69176,3 milhares de toneladas. Deste valor correspondem à produção da União Europeia 46460,6 milhares de toneladas, dos quais mais de 96 % foram produzidos pelos três maiores produtores europeus de azeitona: Espanha, Itália e Grécia. Neste plano económico cabe a Portugal uma quota de mercado europeu de cerca de 2,46 %, sendo que os últimos dados disponíveis apontam para uma produção de 1143,3 mil toneladas em cerca de 1 516 261 hectares (FAOSTAT, 2010).

Em termos de países consumidores de azeite, o principal mercado está focado, para além de na maioria dos países produtores, no plano europeu. O consumo no continente europeu ronda os 67,8 % da percentagem mundial sendo que o restante consumo está distribuído pelo resto do Mundo, com novos nichos económicos nos Estados Unidos da América e Canadá (FAOSTAT, 2010).

Em termos de competitividade o azeite rivaliza diminutamente com os restantes óleos, representando apenas 3,5 % de mercado dos óleos alimentares, devido aos seus elevados custos de produção. Quando comparado com outros óleos, sejam de obtenção primária ou de obtenção secundária, estes são de cultivo anual, com altos rendimentos e com elevado grau de mecanização, por oposição ao azeite, de cultivo e produção laboriosa e longa, resultando num azeite, independentemente da sua qualidade, com valor acrescentado (Boskou, 2006).

*1.3. Normas de comercialização*

A qualidade do azeite obtido depende de vários fatores como a variedade de oliveira, o estado de maturação, a colheita e transporte, bem como o armazenamento e o modo de extração do azeite das azeitonas.

O primeiro passo no processo de obtenção de azeite será a determinação do momento ideal de maturidade das azeitonas, quando o teor de óleo é máximo, normalmente quando a coloração do fruto é verde/amarela ou preta/púrpura, sendo assim colhidas as azeitonas inteiras e sãs. Após a colheita (manual ou mecanizada), segue-se o transporte que deve ser feito em caixas plásticas abertas e com orifícios de modo a permitir que o ar circule, inibindo a atividade enzimática e eventuais aparecimentos de bolores ou fungos.

O processamento deverá ser feito o quanto antes após a colheita. Caso seja necessário esperar algumas horas as azeitonas devem ser armazenadas em locais bem ventilados, com temperatura inferior a 25°C e com humidade relativa inferior a 75 %. Na fase de limpeza as folhas e galhos são removidos por ação de ar forçado e a lavagem é feita com um fluxo de água.

O passo seguinte é a moenda ou trituração, por ação de moinhos de pedra ou de metal, com produção de uma pasta.

Na fase subsequente a pasta é malaxada para permitir a libertação do óleo dos vacúolos das células, onde a oxidação poderá ser minimizada com recurso a um sistema de injeção de gases inertes. A pasta é posteriormente prensada de modo a separar a fase sólida (bagaço) da fase líquida (azeite e água "vegetal"). O azeite é finalmente separado da água por centrifugação ou percolação (filtração seletiva).

O último passo de processamento é o armazenamento, que deverá ser feito em contentores inox de fundo cónico, totalmente cheios e/ou com gás inerte no espaço de cabeça, mantidos ao abrigo da luz e a temperaturas inferiores a 25°C a fim de evitar a oxidação do azeite obtido.

A forma como o azeite é obtido e eventuais tratamentos ao mesmo, permitem a sua classificação e denominação pela Legislação em vigor Regulamento (CE) n.º 1513/2001 do Conselho, de 23 de Julho de 2001, que altera o Regulamento 136/66/CEE, bem como o Regulamento (CE) n.º 1638/98, da seguinte forma:

*Azeites Virgens*

Azeites obtidos a partir do fruto de oliveira unicamente por processos mecânicos ou outros processos físicos – em condições que não alterem o azeite – e que não tenham sofrido outros tratamentos para além da lavagem, da decantação, da centrifugação e filtração, com exclusão dos azeites obtidos por solventes, com adjuvantes de ação química ou bioquímica ou por processos de reesterificação e de qualquer mistura com óleos de outra natureza.

Estes azeites são classificados e descritos do seguinte modo:

*- Azeite Virgem Extra*

Azeite virgem com uma acidez livre, expressa em ácido oleico, não superior a 0,8 g por 100 g e com as outras características conformes com as previstas para esta categoria.

*- Azeite Virgem*

Azeite virgem com uma acidez livre, expressa em ácido oleico, não superior a 2 g por 100 g e com as outras características conformes com as previstas para esta categoria.

*- Azeite lampante*

Azeite virgem com uma acidez livre, expressa em ácido oleico, superior a 2 g por 100 g e/ou com as outras características conformes com as previstas para esta categoria.

*- Azeite Refinado*

Azeite obtido por refinação de azeite virgem, com uma acidez livre expressa em ácido oleico não superior a 0,3 g por 100 g e com as outras características conformes com as previstas para esta categoria.

*- Azeite – Contém Exclusivamente Azeite Refinado e Azeite Virgem*

Azeite constituído por loteamento de azeite refinado e azeite virgem, com exclusão do azeite lampante, com uma acidez livre expressa em ácido oleico não superior a 1 g por 100 g e com as outras características conformes com as previstas para esta categoria.

É também de referir que, o Regulamento (CE) n.° 1019/2002 da Comissão de 13 de Junho de 2002 vem definir regras de normalização de rotulagem, sendo de destacar o seu Art.° 5°, alínea d), onde determina que a menção à acidez ou acidez máxima deve fazer-se sem destaque e acompanhar as restantes informações de qualidade do azeite em questão, a fim de evitar a indução do comprador em erro relativamente às características do azeite, como sendo características especiais que, no entanto, serão comuns à maior parte dos azeites.

## *2. Constituição química do azeite*

Os componentes do azeite distribuem-se essencialmente por duas frações: a fração saponificável, constituída por ácidos gordos esterificados sob a forma de triacilgliceróis (TAGs) que perfazem cerca de 99 % do azeite, e uma fração insaponificável, que contém, compostos fenólicos, esteróis, vitaminas, fosfolípidos e pigmentos (Boskou, 2006; Gigon *et al.*, 2010; Cicerale *et al.*, 2010).

Tabela I – *Principais constituintes do azeite (adaptado de Gigon et al., 2010 e \*Boskou, 2006)*

| Família de Constituintes | Constituintes | Teor (%) |
|---|---|---|
| Ácidos Gordos (99 %) | Ácido Palmítico (C16:0) | 7,5 – 20,0 |
| Ácidos Gordos Saturados (14,8 %) | Ácido Palmitóleico (C16:1) | 0,3 – 3,5 |
| | Ácido Esteárico (C18:0) | 0,5 – 5,0 |
| Ácidos Gordos Monoinsaturados (76,6 %) | Ácido Óleico (C18:1) | 55,0 – 83,0 |
| | Ácido Araquídico (C20:0) | ≤ 0,6; ≤ 0,7* |
| | Ácido Eicosenóico (C20:1) | 0,5; ≤ 0,4* |
| | Ácido Tetracosanóico (C24:0) | ≤ 0,5; ≤ 0,2* |
| Ácidos Gordos Polinsaturados (8,6 %) | Ácido Linolénico (C18:3) | 0,9; ≤ 1,0* |
| | Ácido Linoléico (C18:2) | 3,5 – 21,0 |
| Outras Substâncias (1 %) | Insaponificável, Esqualenos | - |
| Fitoesteróis | $\beta$-sistosterol, Campesterol, Stigmasterol | - |
| Vitaminas | Vitamina E (Tocoferóis), Vitaminas A e K | - |
| Polifenóis | Secoiridóides: Oleuropeína, Dimetileuropeína, Ligstrósido | - |
| | Hidroxitirosol, Tirosol | - |
| Lignanos | Acetoxipinoresinol, Pinoresinol | - |
| Triterpenos | Ácido Oleanólico, Eritrodiol | - |

## 2.1. Ácidos Gordos

Os principais constituintes do azeite são os ácidos gordos esterificados sob a forma de TAGs e em menor quantidade os ácidos gordos sob a forma livre. Existem diversos tipos de ácidos gordos, classificados pelo grau de saturação: as gorduras monoinsaturadas (MUFA), as gorduras polinsaturadas (PUFA) e as gorduras saturadas (SFA).

Destes ácidos gordos do azeite distinguem-se o palmítico, palmitóleico, esteárico, oleico, linoléico e linolénico, que variam na proporção dependendo da amostra, local de produção, clima, variedade de azeitona e grau de maturação do fruto. Na Tabela II, sinopse da tabela anterior é possível aferir a proporção dos ácidos gordos mais relevantes presentes no azeite.

*Tabela II* – Principais ácidos gordos do azeite (adaptado de Gigon *et al.*, 2010 e *Boskou, 2006)

| Ácido Gordo | Fórmula Química | Teor (%) |
|---|---|---|
| Palmítico (C16:0) | $CH_3(CH_2)_{14}COOH$ | 7,5 – 20,0 |
| Palmitóleico (C16:1) | $CH_3(CH_2)_5CH=CH(CH_2)_7COOH$ | 0,3 – 3,5 |
| Esteárico (C18:0) | $CH_3(CH_2)_{16}COOH$ | 0,5 – 5,0 |
| Óleico (C18:1) | $CH_3(CH_2)_7CH=CH(CH_2)_7COOH$ | 55,0 – 83,0 |
| Linoléico (C18:2) | $CH_3(CH_2)_4(CH=CHCH_2)_2(CH_2)_6COOH$ | 3,5 – 21,0 |
| Linolénico (C18:3) | $CH_3CH_2(CH=CHCH_2)_3(CH_2)_6COOH$ | 0,9; ≤ 1,0* |

Como é percetível na interpretação da Tabela II, o ácido óleico é o que se encontra em maior proporção de todos os ácidos gordos, sendo este portanto, o ácido gordo de referência

– sob a forma livre - para a determinação do grau de acidez, expresso em percentagem, conforme referido na Secção 1.3. – Normas de Comercialização.

*2.2. Antioxidantes*

O envelhecimento e morte celular ocorrem no corpo humano pelos mais variados motivos, desde os efeitos da poluição até a motivos de origem endógena, tais como a respiração celular ou mesmo devido ao próprio metabolismo.

No entanto, existem moléculas que devido às suas propriedades e características químicas, permitem abrandar e mesmo inibir tais comportamentos oxidativos, sendo designadas por antioxidantes. Estas moléculas, quando inseridas num meio lipídico, como uma célula ou mesmo numa garrafa de azeite, mesmo em baixas concentrações, possuem um efeito antioxidante, impedindo ou retardando a sua oxidação por inibição dos radicais livres, o que permite classificá-las em dois tipos, respetivamente:

- *Antioxidantes Preventivos*

Reduzem a velocidade do processo de iniciação da oxidação, impedindo ou retardando a formação de radicais livres, desativando os iões dos metais que catalisam a sua decomposição.

- *Antioxidantes "Chain-Breaking"*

São normalmente compostos fenólicos substituídos que bloqueiam a propagação da cadeia, dado que atuam como antioxidantes primários, pela doação de um radical hidrogénio ao radical peróxilo gerado pela oxidação lipídica, formando derivados estáveis (Servili *et al.*, 2009).

No azeite estão presentes diversas substâncias que podem atuar como antioxidantes, entre as quais os carotenóides, compostos fenólicos lipofílicos (tocoferóis) – comuns a outros óleos vegetais – e compostos fenólicos hidrofílicos (álcoois e ácidos fenólicos, flavonóides, lignanos e secoiridóides) - encontrados exclusivamente no azeite (Sánchez-Moreno *et al.*, 1998; Servili *et al.*, 2004).

*2.2.1. Carotenóides*

Os carotenóides são uma família de pigmentos lipossolúveis naturais, de cor amarela, laranja ou vermelha e sensíveis à luz, oxigénio e calor. São normalmente tetraterpenos (Figura 1) (MacDougall, 2002) e estão divididos em duas categorias: os carotenos – constituídos apenas por átomos de carbono e hidrogénio – e as xantofilas – que contêm pelo menos um átomo de oxigénio.

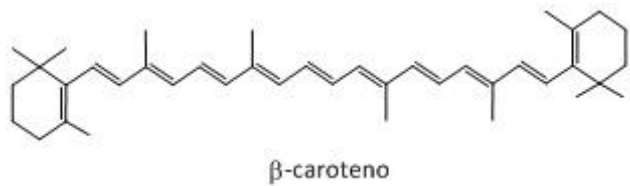

Figura 1 – Exemplo de um caroteno.

A alta temperatura e na ausência de oxigénio, ocorre a isomerização de algumas ligações duplas *trans* a *cis* e ocorre uma consequente diminuição na intensidade da cor. A descoloração também pode ocorrer por ação dos hidroperóxidos, via radicais peróxilo e por ação da lipoxigenase tipo II (Freitas, 2006/2007).

Estas moléculas têm um papel importante nos alimentos e na alimentação pois funcionam como antioxidantes e contribuem para a prevenção de diversas doenças.

Os carotenóides maioritários presentes no azeite são a luteína e o *β*-caroteno (pró-vitamina A), mas poderão existir também diversas (Figura 2) xantofilas: violaxantina, neoxantina, luteoxantina, anteraxantina, mutatoxantina e *β*-criptoxantina (Boskou, 2006).

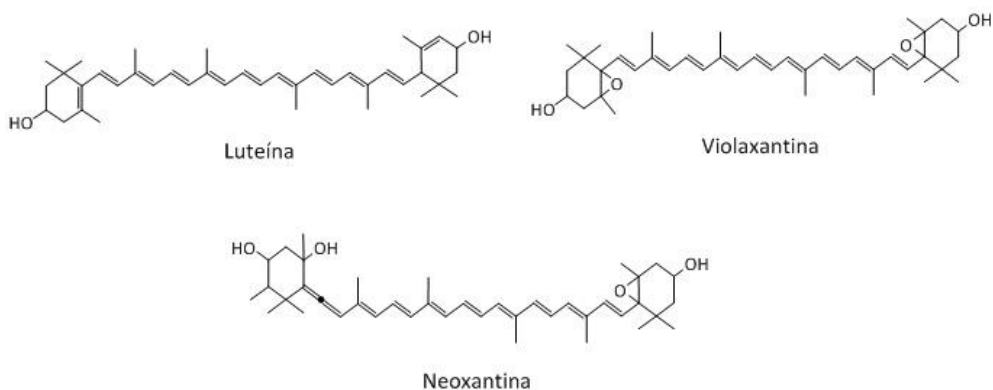

Figura 2 – Exemplos de diferentes xantofilas presentes no azeite.

### 2.2.2. Compostos Fenólicos

Na composição do azeite encontram-se os compostos fenólicos, responsáveis não só pela elevada estabilidade do azeite, como também pelas propriedades biológicas que lhe assistem (Servili *et al.*, 2009).

Estes são diversificados em termos estruturais, mas possuem normalmente um grupo fenólico. Os compostos fenólicos são, ocasional e erradamente, designados por polifenóis, pois nem todos os compostos fenólicos do azeite são polihidroxilados, como por exemplo o tirosol – Figura 3.

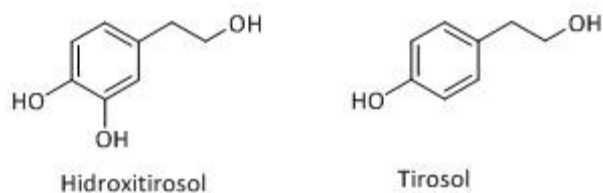

Figura 3 – Estrutura química de um polifenol *per se* e de um composto fenólico.

A atividade antioxidante também está diretamente relacionada com a sua estrutura e com a sua habilidade em transferir o protão do grupo hidroxilo do anel aromático aos radicais livres. Estudos indicam que existe uma maior atividade antioxidante em compostos polihidroxisubstituídos, e preferencialmente nos orto e para substituídos (Tripoli *et al.*, 2005).

Assim, estes compostos são classificados em duas categorias:

- Lipofílicos, como os tocoferóis e tocotrienóis (comuns a todos os óleos vegetais);

- Hidrofílicos, como os álcoois e ácidos fenólicos, flavonóides, lignanos e secoiridóides (exclusivos do azeite).

*Compostos Fenólicos Lipofílicos*

A fração lipofílica dos compostos fenólicos é constituída maioritariamente por tocoferóis (Vitamina E) e também por tocotrienóis (Figura 4). O α-tocoferol representa cerca de 90 % dos tocoferóis do azeite, sendo a restante percentagem constituída pelos diferentes homólogos.

Todos os tocoferóis possuem uma estrutura cíclica de cromano, com o substituinte hidroxilo do anel em posição para relativamente à cadeia alifática formada por 4 anéis isoprénicos.

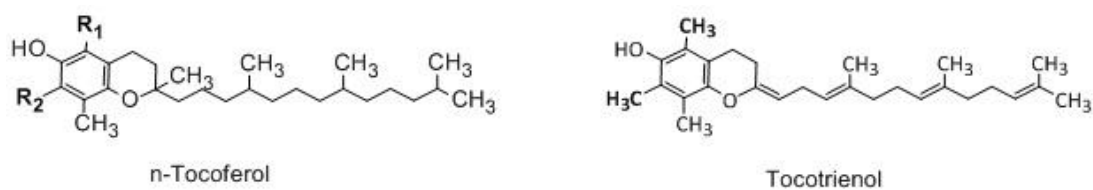

Figura 4 – Comparação entre a estrutura química dos tocoferóis e do tocotrienol, com n= α, β , γ ou σ; onde α: R1= R2= CH3; β: R1= CH3, R2= H; γ: R1=H, R2= CH3; σ: R1= R2= H

Contribuem para a estabilidade oxidativa dos óleos vegetais onde se encontram pois inibem a foto-oxidação por desativação do oxigénio singleto (Boskou, 2006).

Vários estudos demonstram que a atividade dos tocoferóis depende do meio em que estão inseridos e da presença de metais, uma vez que atuam como pró-oxidantes na presença de cobre e ferro (Paiva-Martins *et al.*, 2005; Paiva-Martins *et al.*, 2006).

*Compostos Fenólicos Hidrofílicos*

A fração fenólica hidrófila do azeite é constituída pelos antioxidantes mais abundantes no azeite, dividindo-se, como anteriormente mencionado, em 5 categorias: álcoois e ácidos fenólicos, flavonóides, lignanos e secoiridóides (Tabela III e Figuras 5 a 9).

Tabela III – Composição fenólica do azeite virgem (adaptado de Servili *et al.*, 2009)

| Categoria | Composto Fenólico | Categoria | Composto Fenólico |
|---|---|---|---|
| Ácidos Fenólicos | vanílico | Álcoois Fenólicos | (3,4-dihidroxifenil)etanol |
| | siríngico | | (p-hidroxifenil)etanol |
| | *p*-cumárico | | glicósido do (3,4-dihidroxifenil)etanol |
| | *o*-cumárico | Flavonóides | apigenina |
| | gálico | | luteolina |
| | cafeico | Lignanos | (+)-acetoxipinoresinol |
| | protocatequoico | | (+)-pinoresinol |
| | *p*-hidroxibenzoico | | |
| | ferúlico | | |
| | Cinâmico | | |
| | 4-(acetoxietil)-1, 2-dihidroibenzeno | | |
| | benzóico | | |
| Secoiridóides | 3,4-DHPEA-EDA | | |
| | *p*-HPEA-EDA | | |
| | 3,4-DHPEA-EA | | |
| | aglicona do ligstrósido | | |
| | oleuropeína | | |
| | derivado da p-HPEA | | |
| | forma dialdeídica da aglicona da oleuropeina | | |
| | forma dialdeídica da aglicona do ligstrósido | | |

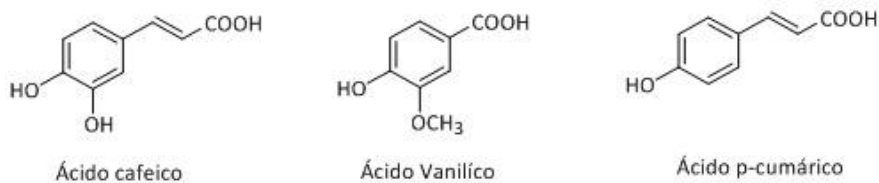

Figura 5 – Estrutura química de alguns ácidos fenólicos.

Em óleos frescos, a concentração de álcoois fenólicos e respetivos derivados é reduzida, mas após o armazenamento estes valores aumentam devido à hidrólise dos secoiridóides presentes no azeite virgem (Servili e Montedoro, 2002).

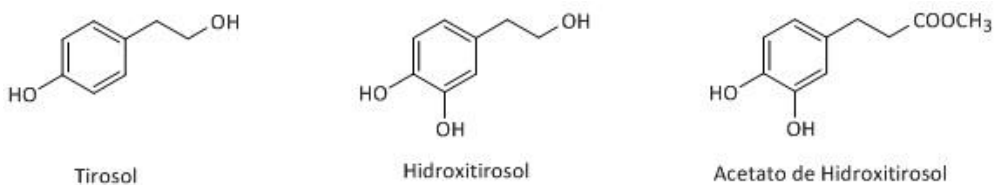

Figura 6 – Estrutura química de alguns exemplos de álcoois fenólicos e seus derivados.

Os flavonóides são antioxidantes presentes naturalmente em várias espécies vegetais, conferindo poder antioxidante às mesmas, devido às suas estruturas fenólicas. São classificados em diversos grupos, com base em critérios estruturais, tais como flavonóis, flavanóis, flavonas, flavanonas, antocianidinas, catequinas e proantocianidinas, sendo que os flavonóides em maior concentração no azeite são a apigenina e a luteolina – Figura 7 – com concentrações que podem variar entre os 0,4 – 2,2 mg.kg-1 e os 0,7 – 7,6 mg.kg-1 (Servili *et al.*, 2009; Freitas, 2006/2007; Carrasco-Pancorbo *et al.*, 2006).

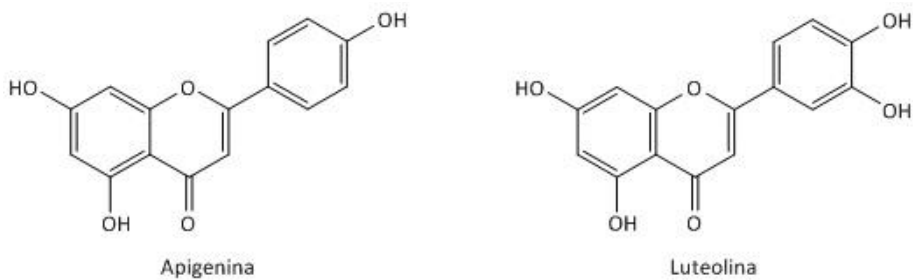

Figura 7 – Estrutura química de alguns flavonoides.

Também estão presentes no azeite os lignanos (+)-1-acetoxipinoresinol e (+)-1-pinoresinol (Figura 8), recentemente isolados e caracterizados, e cujas concentrações podem rondar entre os 25 e 15 mg.kg-1 respetivamente (Servili e Montedoro, 2002).

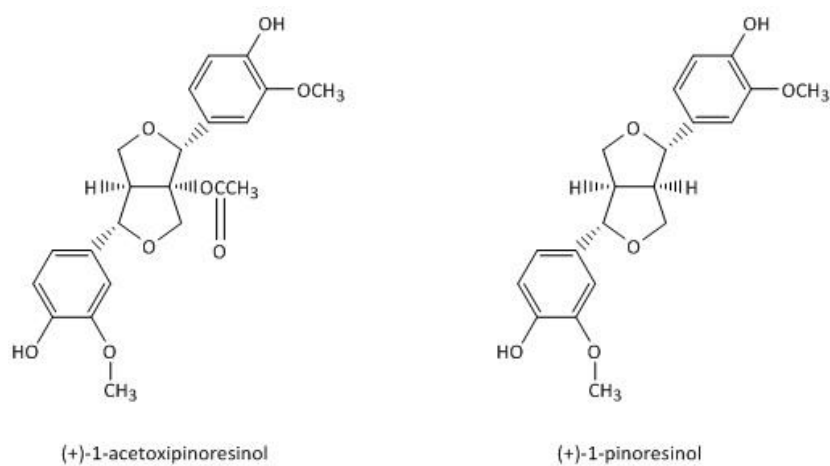

Figura 8 – Estrutura química dos lignanos mais comuns presentes no azeite.

Os secoiridóides são os antioxidantes presentes em maior quantidade no azeite virgem e, ao contrário dos outros compostos fenólicos já referidos, são exclusivos da família *Olearaceae*, onde se incluem as oliveiras.

Estes compostos fenólicos têm tido especial atenção por parte da investigação internacional, pois dos compostos polifenólicos com atividade antioxidante, os que se têm mostrado com maior atividade protetora do azeite e maior atividade biológica são o 3,4-DHPEA-EDA e o 3,4-DHPEA-EA (Servili *et al.*, 2009).

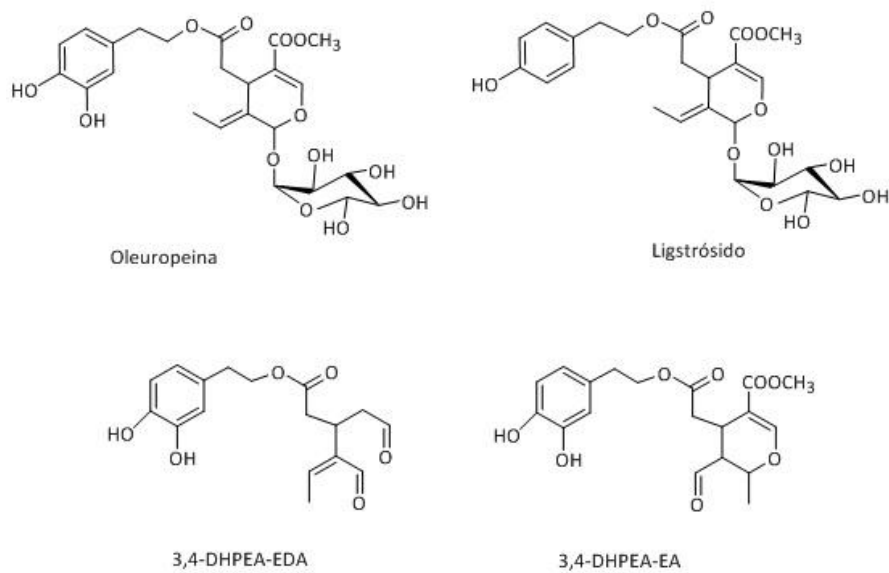

Figura 9 – Estrutura química de alguns dos secoiridóides mais comuns presentes no azeite.

Em termos químicos os secoiridóides são derivados dos iridóides, em que ocorre a clivagem do anel do ciclopentano (Figura 10).

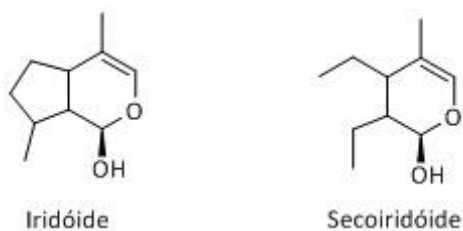

Figura 10 – Estrutura química geral de iridóides e de secoiridóides.

No azeite virgem, os secoiridóides predominantes são os derivados da oleuropeína e do ligstrósido, o 3,4-DHPEA-EDA e o p-HPEA-EDA, e o secoiridóide isómero da oleuropeína aglicona, o 3,4-DHPEA-EA (Servili e Montedoro, 2002; Paiva-Martins *et al.*, 2001). Estes secoiridóides são obtidos por hidrólise enzimática efectuada pelas *β*-glucosidases endógenas do azeite (Servili *et al.*, 2009).

### 3. O Azeite e a Saúde Humana: Qual a relação?

Não é recente nem uma novidade a eventual relação entre a saúde vigorosa dos povos que habitavam a bacia do Mediterrâneo e a sua singular alimentação. A relativa baixa incidência de doenças cardiovasculares, tumorais e neurodegenerativas comparativamente com outros povos despertou e desperta muito interesse na comunidade científica. A dieta típica destes povos mediterrânicos inclui entre outros alimentos, uma fonte de gordura, classificada atualmente como saudável, no azeite.

*3.1. Propriedades Benéficas dos Componentes Minoritários do Azeite*

Um vasto leque de estudos epidemiológicos demonstra o benefício da ingestão do azeite, especialmente quando inserido numa Dieta Mediterrânica (Stark e Madar, 2002; Visioli e Galli, 1995a, 2002; Pérez-Lopéz *et al.*, 2009; Psaltopoulou *et al.*, 2004).

No entanto, os benefícios advêm não só do tipo de gordura que o azeite é - uma gordura essencialmente monoinsaturada – mas também dos componentes minoritários presentes, como os diversos compostos antioxidantes que possui. Desempenha por isso, um papel importante na saúde humana, cujos benefícios são tão vastos e abrangentes como a prevenção de doenças cardiovasculares e de doenças tumorais. É também atribuído um papel quimiopreventivo, com propriedades anti-cancerígenas e anti-inflamatórias a diversos constituintes minoritários do azeite (Servili *et al.*, 2009; Cicerale *et al.*, 2010).

Sendo o azeite uma gordura essencialmente monoinsaturada, quando integrada numa dieta equilibrada, substitui o consumo de gordura saturada e de gordura *trans*, que estão associadas a problemas coronários, elevados valores de LDL, triglicéridos (TG) e colesterol total (TC), entre outros (Cicerale *et al.*, 2010). O azeite é menos suscetível a alterações quando exposto a temperaturas elevadas ou à degradação oxidativa, conservando assim grande parte das propriedades que lhe são atribuídas, sendo essa característica distintiva dos restantes óleos. Alguns autores indicam que esta resiliência é devida não só ao baixo teor em PUFAs mas também aos antioxidantes (Tripoli *et al.*, 2005; Hardwood *et al.*, 2002). Ensaios *in vivo* indicam que a substituição de SFAs por MUFAs permite baixar os valores de TC, LDL e aumentar os de HDL (Covas *et al.*, 2007).

Aos antioxidantes é atribuída a responsabilidade da diminuição do *stress* oxidativo e todas as repercussões nefastas inerentes à presença de radicais livres ao nível celular. Cicerale *et al.* referem a existência de estudos em animais que demonstram que a ingestão de azeite virgem rico em polifenóis melhora o perfil lipídico, bem como o perfil glicídico do sangue, com redução dos valores de TC, TG e LDL, assim como o aumento substancial dos valores de HDL (Gigon *et al.*, 2010; Covas, 2007).

*3.2. Efeitos sobre a Saúde*

Pelo referenciado até aqui, é possível depreender que existem diversos estudos que apontam para um conjunto alargado de benefícios da ingestão de azeite - especialmente quando inserido numa dieta adequada, reforçado por exercício físico regular – e dos seus constituintes, sejam eles os ácidos gordos ou qualquer um dos tipos de antioxidantes. Por estes motivos, os efeitos sobre a saúde por parte do azeite e seus constituintes merecem especial atenção nos pontos seguintes.

*3.2.1. Obesidade*

Os tecidos adiposos do corpo humano servem como camada de protecção em termos climatéricos, funcionando também como reserva energética em períodos em que a ingestão de energia é reduzida ou impossível.

No entanto, com o grau de evolução das sociedades modernas, existe uma tendência de consumismo excessivo a todos os níveis, e em especial a nível alimentar. A sobrealimentação é um problema "urbano" e induzido pelo marketing dos *mass media*, mas também tem explicações fisiológicas como a necessidade de saciedade, levando a comer em maiores quantidades. Não obstante, é um problema de saúde pública que os Governos tentam agora subjugar, quer pelos prejuízos diretos para a saúde individual, quer para os prejuízos a longo prazo para o Estado.

A forma de cálculo do índice de massa corporal (IMC) é a fórmula mais comum de cálculo para determinar o grau de obesidade ou de ausência desta, e é feito da seguinte forma:

Tabela IV - Classificações de Índice de Massa Corporal (Fonte: OMS, Global Database on BMI)

| | | Classificação | IMC (kg/m$^2$) Intervalos de classificação |
|---|---|---|---|
| Abaixo do Peso Ideal | <18.50 | Magreza Grave | < 16,00 |
| | | Magreza Moderada | 16,00 – 16,99 |
| | | Magreza Leve | 17,00 – 18,49 |
| Peso Ideal | 18.50 - 24.99 | Peso Ideal | 18,50 – 24,99 |
| Acima do Peso Ideal | ≥25.00 | Pré-Obesidade | 25,00 – 29,99 |
| Obesidade | ≥30.00 | Obesidade Classe I | 30,00 – 34,99 |
| | | Obesidade Classe II | 35,00 – 39,99 |
| | | Obesidade Classe III | ≥ 40,00 |

A sobrealimentação causa o acumular de energia não necessária/consumida nos tecidos adiposos sob a forma de TAGs, levando assim à obesidade e aos problemas inerentes a essa nova condição: dificuldades de locomoção, dificuldades respiratórias, problemas ósseos e cardiovasculares.

O efeito do consumo de gorduras essencialmente monoinsaturadas, sob a forma de azeite, enquadrada numa dieta Mediterrânica, foi estudado por Bes-Rastrollo *et al.*, e revelou que um elevado consumo de azeite não está associado a um maior ganho de massa corporal, ou a um aumento do risco de desenvolver obesidade no contexto de uma dieta Mediterrânica, resultados em linha com outros estudos efetuados (Serra-Majem *et al.*, 2003). O consumo de azeite está também ligado a um melhoramento do perfil glicídico e insulínico (Gigon *et al.*, 2010), bem como a um potencial de saciedade maior induzido pela transformação do ácido oleico no intestino, numa hormona responsável pelo aumento de saciedade – a oleoiletanolamida (OEA) - reduzindo assim a necessidade de comer em excesso para obter o mesmo efeito. Schwartz e seus colaboradores reportam a transformação do ácido oleico

ingerido – no qual o circulante não demonstrou o mesmo comportamento – em OEA, um mensageiro lipídico que se liga a recetores-$\alpha$ ativados por proliferador de peroxíssomas (PPAR-$\alpha$) e que não se verifica em cobaias geneticamente modificadas para a ausência do transportador de ácido gordos CD36. Este facto sugere que a ativação da mobilização do OEA do intestino delgado, possibilitada pelo transportador CD36, serve de sensor, relacionando a ingestão lipídica com a saciedade (Schwartz *et al.*, 2008). O interesse acerca desta hormona é crescente na comunidade científica e na investigação relacionada com a obesidade pois diferentes estudos referem também que, a OEA exógena regula o tempo entre refeições (Oveisi *et al.*, 2004), diferenciando-a de outros péptidos de origem intestinal, como a colecistoquinina (CCK) que regula, não a frequência mas o tamanho das refeições (Moran, 2006).

*3.2.2. Stress Oxidativo*

Os efeitos do *stress* oxidativo (desequilíbrio no balanço produção/remoção de espécies reativas de oxigénio – ROS – consideradas subprodutos resultantes da respiração aeróbia) são vários e levam a repercussões em diversos componentes do organismo, tais como danos no ADN, nos eritrócitos, com influência direta em doenças como aterosclerose e o surgimento de problemas cardiovasculares (Servili *et al.*, 2009; Cicerale *et al.*, 2010).

*3.2.2.1. Aterosclerose e Doenças Cardiovasculares*

Tem-se verificado, nos últimos anos, um aumento de interesse na influência do azeite e seus constituintes na modificação oxidativa das LDL (Torre-Carbot *et al.*, 2007; Cicerale *et al.*, 2003; Servili *et al.*, 2009; Tripoli *et al.*, 2005). Crê-se que um passo-chave no desenvolvimento de doenças cardiovasculares é a aterosclerose, caracterizada pelo endurecimento e espessamento das paredes arteriais, devido à acumulação de lípidos, tecidos fibrosos e cálcio nas mesmas (Sies, 1997).

Circulam na corrente sanguínea lípidos (TAGs, fosfolípidos, colesterol esterificado e colesterol livre) que se associam a proteínas – génese da designação lipoproteínas – cuja função é transportar esses mesmos lípidos desde os locais de absorção até aos locais onde se processa a sua degradação e posterior utilização, e respetiva excreção. Existem diferentes tipos de lipoproteínas classificados de acordo com a respetiva densidade e função: quilomicra, lipoproteínas de muito baixa densidade (VLDL), lipoproteínas de baixa densidade (LDL) e lipoproteínas de elevada densidade (HDL).

O "mau colesterol" são cientificamente as lipoproteínas (LDL), que transportam colesterol na corrente sanguínea, do fígado para os tecidos. Os fatores que influenciam os seus níveis no sangue tendem também a afetar a concentração total de colesterol. A oxidação das LDL causa danos nas paredes vasculares, estimulando a sua endocitose pelos macrófagos e a formação de células esponjosas que levarão à formação de ateromas. Fatores necróticos libertados por estas células permitem uma penetração mais fácil de monócitos na íntima que, por endocitose de LDLox levam à formação de placas dentro das paredes arteriais.

Por este motivo, teores elevados de TC e de LDL foram estabelecidos como fatores de risco para o desenvolvimento de aterogénese, a causa principal de problemas cardiovasculares (Cicerale *et al.*, 2010; Covas *et al.*, 2007; Tripoli *et al.*, 2005; Sies, 1997).

O vulgarmente chamado de "bom colesterol" é cientificamente designado por HDL e transporta o excesso de colesterol do local onde se encontra (células dos tecidos ou órgãos periféricos) para o fígado, onde é reutilizado ou excretado pela via biliar através de duas formas possíveis: sob a forma livre, ou sob a forma de ácidos biliares. Por este motivo, crê-se que as HDL são benéficas na saúde humana, atuando como fator de proteção na prevenção de doenças coronárias (Fennema, 1996; Harwood e Yaqoob, 2002) Alguns estudos indicam que a ingestão de azeites virgens com elevado teor de fenóis leva ao aumento de HDL no sangue, em alguns casos com valores de cerca de 5 - 7%, bem como à diminuição dos valores de LDL (Cicerale *et al.*, 2010).

O *stress* oxidativo pode ser indiciado pela presença de determinados marcadores, como isoprostanos-F2, peróxidos lipídicos (LPO), glutationa oxidada (GSSG), e peroxidase glutationa (GSH-Px). O estudo in vivo realizado por Biswas *et al.* em 2005 demonstrou que os isoprostanos-F2 são resultado da peroxidação induzida do ácido araquidónico pelos radicais livres, enquanto os LPO serão um subproduto da oxidação dos ácidos gordos e do esgotamento da GSH que antecede a oxidação lipídica e a aterogénese (Cicerale *et al.*, 2010). Igualmente em 2005, Ruano *et al.* demonstraram os efeitos benéficos da ingestão de azeite com elevado teor em compostos fenólicos em pacientes hipercolestorémicos (com elevado teor de colesterol sanguíneo), através da redução dos níveis de isoprostano-F2, resultados coadunantes com o estudo executado por Visioli *et al.*, 2000a.

Vários polifenóis do azeite têm vindo a mostrar efeitos benefícios em sistemas biológicos – o hidroxitirosol (Figura 3, Secção 2.2.2.) – mostrou em estudos *in vitro*, inibir totalmente a agregação das plaquetas em sangue humano, tendo sido notados comportamentos inibitórios semelhantes em outros antioxidantes do azeite, como a luteolina e a aglicona da oleuropeína (Figura 7 e 9 respetivamente, Secção 2.2.2.) (Servili *et al.*, 2009). No seu estudo *ex vivo* de 1995, Visioli e seus colaboradores sugerem que a oxidação das LDL é inibida por ação dos constituintes do azeite extra-virgem. Ainda noutro estudo, dois fatores pró-coagulantes, o plasminogéneo ativador inibidor-1 (PAI-1) e o fator VII (FVII) viram as suas concentrações reduzidas por ação de azeite virgem de elevado teor polifenólico (Ruano *et al.*, 2007).

Assim o azeite e os seus constituintes poderão apresentar um papel importante na prevenção da oxidação das LDL e dos lípidos, inibindo a aterogénese (Tripoli *et al.*, 2005).

Sob outro ponto de vista na linha de investigação do *stress* oxidativo, existem estudos que indicam outros efeitos nefastos na saúde humana: desde danos aos eritrócitos e ao ADN, até à relação com doenças neurodegenerativas.

Os eritrócitos são células sanguíneas, anucleadas, responsáveis pelo transporte de oxigénio. Possuem uma capacidade reduzida de auto-reparação e sendo células de transporte específico de oxigénio, estão expostas a danos oxidativos sempre que se desenvolve uma situação de *stress* oxidativo. No entanto, possuem antioxidantes endógenos de modo a proteger minimamente a sua integridade e funcionalidade, tais como enzimas antioxidantes, glutationa, tocoferóis e ascorbato. Não obstante, ocorrem constantemente no eritrócito danos parciais ou irreversíveis, marcando a célula para a sua remoção após uma vida de cerca de 120 dias (Paiva-Martins *et al.*, 2009). No entanto, o eritrócito torna-se importante para o estado oxidativo do sangue não só porque é uma fonte de ROS mas também porque é o principal local de regeneração do ácido ascórbico sanguíneo, dependendo esta regeneração da integridade das suas defesas antioxidantes.

Um estudo recente de Paiva-Martins *et al.* (2010) demonstra a interação do 3,4-DHPEA-EDA com os eritrócitos, protegendo-os da hemólise oxidativa iniciada por radicais peróxilo. Também fica subjacente dos resultados obtidos que, a proteção não advém apenas de um mecanismo de "*radical scavenging*" mas possivelmente também de uma interação direta com as membranas celulares, induzindo modificações ao perfil proteico da mesma.

*3.2.2.2. ADN*

Para além da hemólise induzida nos eritrócitos também danos a nível do ADN poderão acontecer e já foram reportados em alguns estudos (Salvini *et al.*, 2006; Silva *et al.*, 2008; Yermilov *et al.*, 1995) que visaram demonstrar o potencial antioxidante dos compostos polifenólicos do azeite na prevenção do dano ou na reparação do mesmo. Muita atenção tem sido dada a esta linha de investigação pois os danos oxidativos no ADN estão na génese de muitos carcinomas (Covas, *et al.*, 2007; Cicerale *et al.*, 2010; deRojas-Walker *et al.*, 1995; Baskin e Salem, 1997). Muito embora os polifenóis tenham propriedades antioxidantes que permite prevenir a oxidação lipídica e proteica, alguns podem agir também como pró-oxidantes, induzindo danos oxidativos em células, sendo por isso citotóxicos. Esta citotoxicidade tem-se mostrado em alguns casos seletiva, tendo-se verificado ser superior em células cancerígenas. Assim, certos antioxidantes poderão desempenhar um duplo papel na mutagénese e carcinogénese, dependendo a sua atuação como anti ou pró-oxidantes do estado redox do ambiente biológico em que está inserido (Baskin e Salem, 1997).

O radical hidroxilo é o mais comum dos radicais capazes de induzir danos no ADN, pois possui elevada eletrofilicidade e alta reatividade termocinética. Devido às suas propriedades, pode atacar o ADN por duas vias possíveis: abstração de protões da desoxirribose ou pela adição às ligações π do ADN. No primeiro caso, a abstração de átomos de hidrogénio da desoxirribose pode levar à cisão da estrutura açúcar-fosfato numa das cadeias de ADN helicoidal e cabe a certas enzimas reparadoras retificarem os danos. No entanto, quando o ataque de radicais hidroxilo é excessivo, os danos podem acontecer nas duas cadeias de ADN, o que poderá levar à cisão da estrutura de ADN resultando na morte celular. No segundo caso, a adição às ligações π das bases leva, entre outras, à formação de 8-hidroxi-2'-desoxiguanosina (8OH2dG), uma das lesões mais mutagénicas conhecidas. Esta lesão pode levar a alterações na velocidade de replicação do ADN, resultando em mutações, hidrólise ou cisão das cadeias (Baskin e Salem, 1997).

Estudos demonstraram que o consumo de azeite com elevado teor em polifenóis consegue reduzir em 30% (Salvini *et al.*, 2006) os danos oxidativos no ADN, bem como diminuir a concentração de marcadores específicos da ocorrência de danos oxidativos no ADN na urina (Machowetz *et al.*, 2007; Weinbrenner *et al.*, 2004). Outros estudos revelam que três polifenóis específicos, entre os quais um flavonóide característico do azeite, a luteolina (Figura 7), têm efeitos protetores contra os danos oxidativos ao ADN (Silva *et al.*, 2008).

*3.2.2.3. Cancro*

São diversos os estudos epidemiológicos que demonstraram uma associação entre o consumo de azeite e o risco reduzido de contrair diferentes tipos de cancro – da mama, da próstata, dos pulmões, da laringe, dos ovários e do cólon (Servili *et al.*, 2009; Filik e Ozyilkan, 2003; Covas *et al.*, 2007).

Colomer e Menéndez, no seu estudo de 2006, referem a capacidade do ácido óleico regular diretamente a expressão de certos oncogénios, com papel crucial na transformação maligna, tumorogénese, metástase e insucesso nos tratamentos anti-cancerígenos, e responsáveis por cerca de 20 % dos cancros da mama e agressividade dos mesmos. Mais recentemente, outros estudos revelaram que o ácido oleico atua sinergeticamente com uma certa terapia anti-cancerígena, inibindo o crescimento celular de células cancerosas mamárias MCF-7 e SKBR3 (Menéndez *et al.*, 2007; Menéndez *et al.*, 2008; Menéndez *et al.*, 2009).

Para além do ácido oleico, alguns fenóis, nomeadamente a oleuropeína e o hidroxitirosol, demonstraram a capacidade de reduzir a taxa de proliferação celular (Fabiani *et al.*, 2002; Fabiani *et al.*, 2006) e induzir a apoptose nas células MCF-7 (Han *et al.*, 2009). Mais recentemente, o hidroxitirosol demonstrou também inibir a proliferação do adenocarcinoma humano, através da forte inibição da fosforilação das quinases reguladas por sinais extracelulares (ERK1/2) (Corona *et al.*, 2009).

*3.2.2.4. Doenças Neuro-Degenerativas*

Ainda relacionada com os efeitos do *stress* oxidativo existem as doenças neuro-degenerativas como a doença de Alzheimer.

A doença de Alzheimer é uma doença de presença crescente na sociedade atual, incurável, degenerativa e terminal. É caracterizada por irritabilidade, confusão, variações bruscas de humor e perda de memória, culminando em perda funcional e morte do paciente. O desenvolvimento da doença não é ainda bem compreendido, tal como não existe uma cura ou um método preventivo e os tratamentos atuais apenas atenuam os sintomas. Duas novas abordagens poderão permitir obter informações cruciais. Numa linha de estudo, está a ser analisada uma proteína – designada por Tau – que está ligada à construção e estabilização de microtubulos e que, em certos casos se agrega sob a forma de emaranhados neurofibrilares. Outra linha de estudo incide sobre os oligómeros beta-amiloides (A$\beta$) – também chamados por ADDL (ligandos amilóides difusíveis beta-derivados) – poderem estar relacionados com desenvolvimento da doença de Alzheimer por se ligarem aos neurónios, causando alterações eléctricas ao nível neuronal e ao nível sináptico, causando problemas de memória. Dois estudos diferentes, ambos de 2009, incidiram sobre estas duas vias: Li *et al.* revelaram que o Ty-EDA inibe a agregação da proteína Tau, enquanto Pitt e seus colaboradores demonstraram que o Ty-EDA tem a capacidade de alterar o estado da oligomerização dos ADDL protegendo os neurónios dos respetivos efeitos nefastos.

Outra doença do foro degenerativo cuja relação com o *stress* oxidativo parece ser uma realidade é a doença de Parkinson (Castellani *et al.*, 1996; Jenner, 2003). É uma doença perturbadora do sistema nervoso central (SNC), do tipo crónico e progressivo, que prejudica as capacidades motoras do indivíduo, em como a capacidade comunicativa.

*3.2.2.5. Inflamação e Dor*

Após a demonstração das propriedades benéficas do azeite incluído numa dieta do tipo Mediterrânico, as atenções dos investigadores centraram-se primeiramente nos seus principais constituintes e depois nos minoritários. Estes últimos revelaram propriedades antioxidantes, capazes de reduzir e mesmo inibir o *stress* oxidativo de eritrócitos, das cadeias de ADN e deste modo ter uma influência direta sobre doenças cardiovasculares.

Recentemente foram publicados alguns estudos que revelam efeitos sobre modulação da inflamação e da atividade endotelial (Covas *et al.*, 2007), como a diminuição da concentração de marcadores biológicos da inflamação, como a interleuquina (Il-6) e proteína C reativa (PCR) (Alemany *et al.*, 2010), a diminuição das concentrações plasmáticas de tromboxano B2 (TXB2) e leucotrieno B4 (LTB4) (Bogani *et al.*, 2007; Visioli *et al.*, 2005; Léger *et al.*, 2005). Também foram reportados efeitos sobre a capacidade de modular a expressão de certas metaloproteases (MMP-9), pela ação da via do fator nuclear kappa B (NF-kB) (Verma *et al.*, 1995) – e a diminuição da concentração de alguns mediadores de inflamação como o fator de necrose tumoral alfa (TNF-α), a interleuquina-1-β (IL-1-b) e a prostaglandina E2 (PGE2) (Miles *et al.*, 2005).

Para além da diminuição dos processos inflamatórios, diversos estudos indicam que a atividade dos polifenóis inibem a formação de ciclooxigenases (COX) – enzima responsáveis pela formação de mediadores, como as prostaglandinas e os tromboxanos (Beauchamp *et al.*, 2005). A COX existe sob três formas, correspondentes a três isoenzimas, designadas por COX-1 (considerada uma enzima constitutiva e presente na maioria da células dos mamíferos, com funções fisiológicas), COX-2 (uma enzima induzida, encontrada em macrófagos ativados e em outras células em locais de inflamação) e COX-3 (enzima cuja função exata é desconhecida mas se crê estar relacionada com o SNC). A inibição da COX leva à diminuição sintomática da dor e inflamação, sendo este o modo de atuação de medicamentos não-esteróides como o ibuprofeno ou a aspirina.

Observou-se uma actividade anti-inflamatória semelhante à do ibuprofeno ou à da aspirina por parte do Ty-EDA (Figura 11) (Beauchamp *et al.*, 2005), obtida pela inibição da COX-1 e COX-2. (Romero *et al.*, 2007; Cicerale *et al.*, 2010; Covas *et al.*, 2007; Servili *et al.*, 2009; Tripoli *et al.*, 2005; Obied *et al.*, 2008).

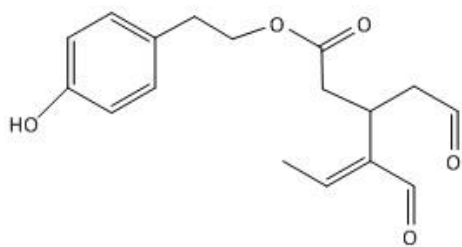

Figura 11 – Estrutura química do Ty-EDA.

*3.2.2.6. Ação Anti-Microbiana*

Outros trabalhos afirmam também que, certos compostos fenólicos apresentam atividade antibacteriana e/ou bactericida contra numerosos patogénios, como a *Listeria monocytogenes*, a *Staphylococcus aureus*, a *Shigella sonnei* e a *Salmonella enterica* (Romero *et al.*, 2007). Romero *et al.* reportaram também atividade anti-*Helicobacter pylori* (bactéria do revestimento mucoso do estômago, responsável por úlceras pépticas, gastrites e cancros estomacais) por parte do Ty-EDA, em quantidades tão reduzidas quanto 1,5 µg.mL-1. A atividade anti-microbial de outros compostos fenólicos foi também demonstrada, nomeadamente para a oleuropeína, o hidroxitirosol e tirosol (Medina *et al.*, 2006; Besignano *et al.*, 1999; Tripoli *et al.*, 2005).

*3.3. Biodisponibilidade dos compostos fenólicos do azeite*

As propriedades biológicas dos compostos fenólicos do azeite dependem da extensão da absorção e consequente metabolismo, pelo qual são diversos os estudos *in vivo* efetuados a fim de investigar a biodisponibilidade dos fenóis do azeite (Visioli *et al.*, 2000b; Vissers *et al.*, 2002; Miró-Casas *et al.*, 2001; Miró-Casas *et al.*, 2003a, Miró-Casas *et al.*, 2003b).

Nestes estudos determinaram-se os níveis de tirosol na urina e sangue, cujos valores aumentam após a ingestão de azeite, verificando-se que estes são excretados essencialmente sob a forma de *O*-glucoronídeos, tanto dos compostos parentais como do ácido e álcool homovanílico (Visioli *et al.*, 2000b; Visioli *et al.*, 2003; D'Angelo *et al.*, 2001; Caruso *et al.*, 2001), demonstrando que nos humanos estes compostos sofrem a ação da catecol-*O*-metiltransferase (COMT) (Tripoli *et al.*, 2005; Caruso *et al.*, 2001). Destes estudos resulta também que a excreção destes compostos estará correlacionada com a dose administrada (Caruso *et al.*, 2001; Vissers *et al.*, 2002; Visioli *et al.*, 2000b). Em 2006, Corona e seus colaboradores, efetuaram um estudo no qual reportaram uma concentração dos glucoronídeos do HyTy e do HVA maior que a administrada, sendo portanto provável que o hidroxitirosol tenha origem também nos secoiridóides.

Mais recentemente, confirmou-se que após a administração de secoiridóides como 3,4-DHPEA-EDA e 3,4-DHPEA-EA, aparecem nos líquidos serosais não só os glucoronídeos dos compostos parentais reduzidos, como os glucoronídeos do HyTy e HVA (Pinto *et al.*, 2010).

*3.3.1. Metabolismo*

Após a entrada de um xenobiótico no organismo, inicia-se a metabolização do mesmo. A metabolização tem como função converter uma molécula capaz de atravessar membranas biológicas, numa que possa ser excretada (pela urina, suor, etc.). Ao longo dos passos metabólicos, a lipofilicidade das moléculas normalmente reduz-se (o que se traduz também no aumento de polaridade), e é este fator que determina se ela vai para a excreção renal ou se prossegue para a metabolização, podendo qualquer um dos passos ativar ou desativar o xenobiótico – o que poderá limitar a ação de fármacos (se desativado) ou funcionalizar um pró-fármaco (se ativado).

O metabolismo de xenobióticos é dividido em duas fases distintas: fase I e fase II, onde cada uma destas comporta diferentes transformações ao xenobiótico.

Após a ingestão do xenobiótico, verifica-se a sua absorção pela corrente sanguínea e entra no sistema porta-hepático e através da veia porta chega ao fígado. Aí, a sua concentração pode ser bastante reduzida antes de ser lançada no sistema circulatório, motivo pelo qual, quando se trata da ingestão de medicamentos, alguns poderão necessitar de ser administrados por vias alternativas à gastro-intestinal, evitando assim a ação hepática, tal como administração sublingual, intravenosa, inalação, entre outras. No entanto, o metabolismo de Fase I e de Fase II não é exclusivo de um determinado órgão, podendo ocorrer em diferentes locais ou órgãos. Assim, a maioria acontece a nível do fígado mas muitas – como a hidrólise de ésteres e amidas, a redução e até mesmo a glucoronoconjugação – poderão ocorrer também a nível da parede intestinal, plasma sanguíneo e outros tecidos.

Relativamente à forma como o metabolismo dos xenobióticos é processado, na fase I, ocorrem essencialmente três tipos de reações: reações oxidativas, redutoras e hidrolíticas. Dentro das reações oxidativas poderão ocorrer hidroxilações aromáticas, epoxidação de alcenos, oxidação de carbonos adjacentes a centros $sp^2$, oxidação de átomos de carbono alifáticos e alicíclicos, oxidação de sistemas C-N, C-O e C-S, oxidação de álcoois e aldeídos bem como outras reações oxidativas. Dentro das reações de carácter redutor poderão ocorrer reduções a grupos carbonilo, de grupos nitro, de grupos azo, de grupos azido, redução de óxidos de aminas terciárias, desalogenação redutiva e reações carboxilativas. Os ésteres e as amidas são suscetíveis à ação das estereases, enquanto os grupos nitro, azo e carbonilo são suscetíveis à ação das reductases.

Enzimas não específicas, tal como as do citocromo $P_{450}$ no fígado, têm a capacidade de introduzir grupos funcionais polares em xenobióticos, aumentando a sua polaridade e tornando os mesmos mais passíveis de sofrerem excreção renal. Outras reações enzimáticas poderão "desmascarar" grupos funcionais já existentes, resultando assim também no aumento de polaridade da molécula.

A fase II do metabolismo ocorre principalmente no fígado, mas vários estudos têm mostrado que o hidroxitirosol sofre extensa O-metilação e glucoronização ao nível do intestino (Corona *et al.*, 2009; Pinto *et al.*, 2010). Nesta fase ocorre a conjugação dos metabolitos resultantes do metabolismo de fase I a pequenas moléculas polares endógenas tais como o ácido glucorónico, sulfato e aminoácidos. Estas alterações nas moléculas podem desativar ainda mais um xenobiótico por alteração direta das suas propriedades. Nesta fase ocorre, normalmente, produção de metabolitos hidrofílicos (por redução da sua lipofilicidade, como referido anteriormente), que são excretados na bílis (se a MM > 300 g.mol-1) ou na urina. Também pode ocorrer intervenção da glutationa, que captura metabolitos de elevada eletrofilicidade antes de modificarem macromoléculas importantes a nível biológico como o ARN, o ADN ou proteínas. As reações de conjugação acontecem principalmente em grupos hidroxilo, carbonilo, amino, tiol e azoto heterocíclico, em reações de glucuronização, acetilação, metilação, entre outras (Silverman, 2004; Patrick, 2005).

Uma outra enzima de papel preponderante no metabolismo é a COMT, que intervém na desativação das catecolaminas. Devido à semelhança estrutural do hidroxitirosol com as catecolaminas, este vai ser metabolizado pela COMT a álcool homovanílico, pela introdução de um grupo metilo na posição 3-hidroxi do anel. O hidroxitirosol é também um metabolito do neurotransmissor dopamina (Figura 12) (Manna *et al.*, 2000).

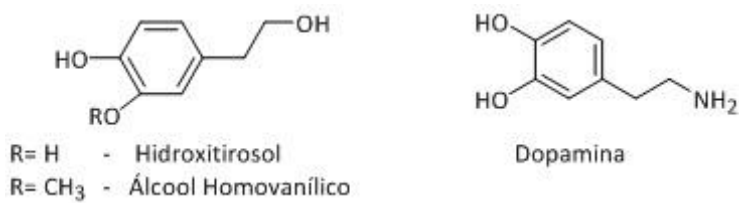

Figura 12 – Estruturas químicas do hidroxitirosol, seu metabolito álcool homovanílico e dopamina.

Relativamente aos metabolitos de compostos fenólicos do azeite, o hidroxitirosol é convertido, por via enzimática, em quatro derivados oxidados e/ou metilados, sendo eles o ácido 3,4-dihidroxifenilacético, o 3,4-dihidroxifenilacetaldeído, bem como o HVA e HVAc – cujos caminhos metabólicos estão representados na Figura 13 (D'Angelo *et al.*, 2001).

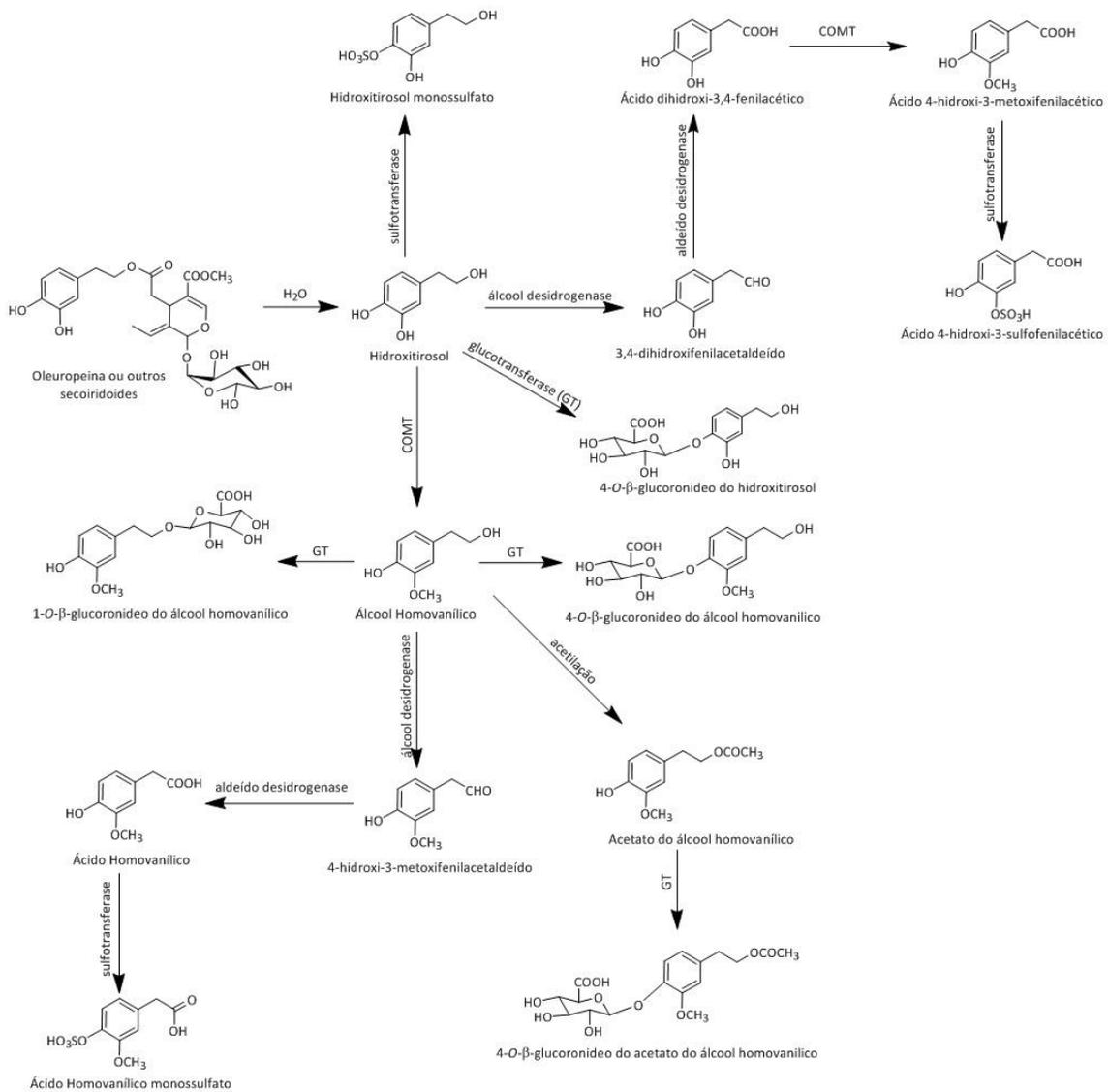

Figura 13 – Caminhos metabólicos possíveis do HyTy - metabolito da oleuropeína - e precursor do HVA (adaptado de Khymenets *et al.*, 2001; D'Angelo *et al.*, 2001; Tuck *et al.*, 2002).

Os produtos de metabolização dos compostos dependem da estrutura química inicial do fenol em causa. No caso do hidroxitirosol, a sua glucuronização biocatalizada no fígado, ocorre nos grupos hidroxilo do anel aromático, onde a posição 4-hidroxi é preferida à 3-hidroxi, sendo esta regiosseletividade comum nos humanos e nos ratos. Relativamente ao tirosol e ao álcool homovanílico, formam-se dois tipos de glucuronídeos: o *O-β*-D-glucuronídeo fenólico e o *O-β*-D-glucuronídeo de álcool alifático, sendo que o glucuronídeo fenólico se forma em maior quantidade comparativamente ao último (Khymenets *et al.*, 2006).

## 4. Síntese de O-Glicósidos

A síntese de glicósidos é uma área de franco crescimento nos últimos anos, devido ao aumento de interesse nas possíveis aplicações em diversas áreas. No entanto, as suas aplicações começam nos estudos científicos que incorporam oligosacarídeos na produção de anticorpos, interações com vírus, substratos para glicosidases e glicosiltransferases, bem como na especificidade de lectina e selectina (proteínas com propriedades de ligação reversível a açúcares), à semelhança do estudo efetuado por Sébastien Marti de 2004 (Ali, 2003).

Existem essencialmente dois modos de conseguir a formação de ligações glicosídicas: o método enzimático e o químico.

### 4.1. Método Enzimático

No primeiro caso, são utilizadas enzimas – tais como as glicosil-transferases – onde os açúcares mono ou difosfato são dadores glicosil, os seus resíduos são os grupos de saída e os açúcares ou outras agliconas são os dadores glicosil. Apesar da elevada especificidade na obtenção de determinadas regio e diasteroseletividades, o elevado custo das enzimas, bem como as dificuldades inerentes ao método reacional e a geração complicada e dispendiosa de dadores glicosil limitam a utilização deste método para oligosacáridos complexos (Ali, 2003).

### 4.2. Método de Síntese Química

No segundo método, a síntese de glicósidos é conseguida através de reações de glicolisação, onde um dador glicosil protegido, com um grupo de saída (GS) adequado no seu carbono anomérico é combinado com um aceitador glicosil total ou parcialmente protegido à excepção de, tipicamente, um grupo hidroxilo livre onde a ligação O-glicósido se irá formar. O dador (açúcar) irá actuar, tipicamente, como um eletrófilo e o aceitador como um nucleófilo, onde o grupo de saída do dador glicosil e os grupos protetores serão os fatores determinantes no rendimento e na seletividade anomérica das reações de glicolisação, dado que uma mistura anomérica $\alpha/\beta$ é formada, como representado na Figura 14 (Ali, 2003).

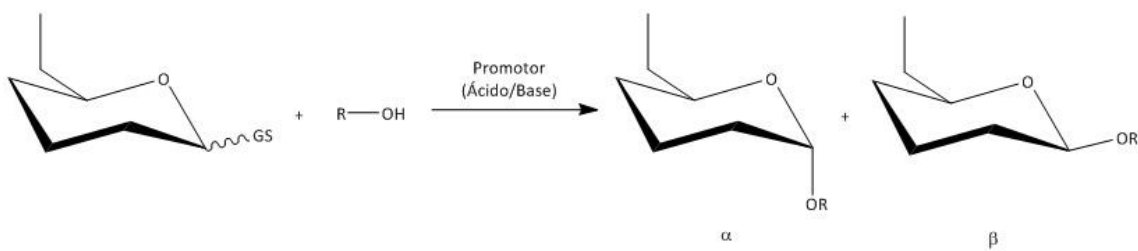

Figura 14 – Tipificação das reações de glicolisação.

*4.2.1. Método dos Fosfitos*

Este é um outro método desenvolvido por Schmidt e seus colaboradores como complemento ao método anterior e tem aplicabilidade em deoxiaçúcares, sendo usado no passo da sililação na síntese de glicósidos do ácido neuramínico (Schmidt *et al.*, 1993 Schmidt, 1998).

A síntese destes glicósidos é feita a partir do oxigénio anomérico desprotegido dos açúcares pela reação com clorofosfitos e com base de Hünig. Os *β*-glicosil fosfitos do ácido neuramínico podem ser ativados com quantidades catalíticas de TMSOTf (Ali, 2003), na reação com ROH para dar os compostos glicosídicos.

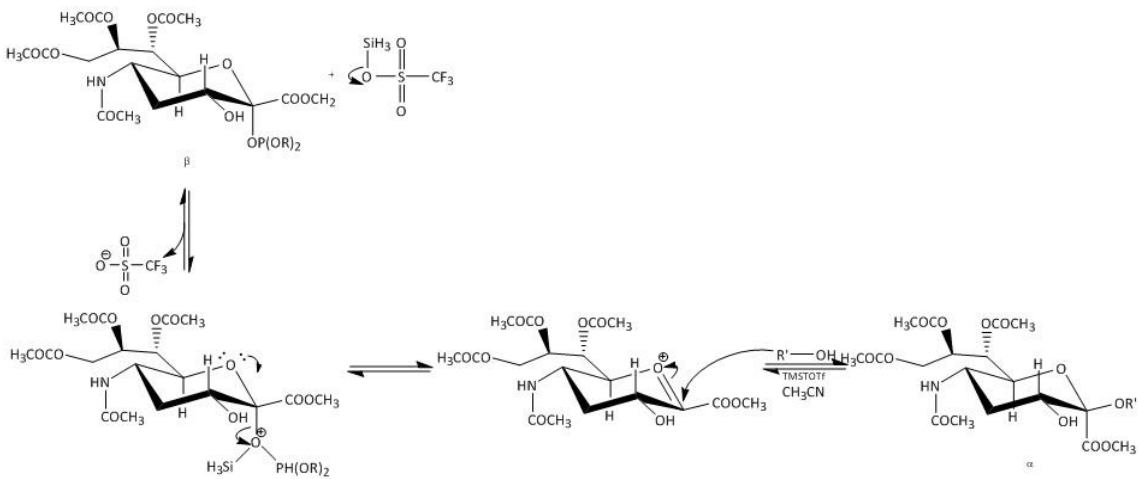

Figura 15 – Mecanismo proposto para a glicolisação via fosfitos.

*4.2.2. Método Intramolecular*

Este método pode, em princípio, superar as questões da ativação e de estereoquímica quando se tenta manter a orientação adequada dos componentes glicosídicos de modo a estes poderem ser forçados a se acoplarem intramolecularmente.

É um método recente mas apresenta um elevado potencial na síntese de ligações açúcar-açúcar de oligosacáridos. Os métodos intramoleculares encontram-se divididos em três categorias: de substituinte funcional, de espaçador rígido e de grupo de saída.

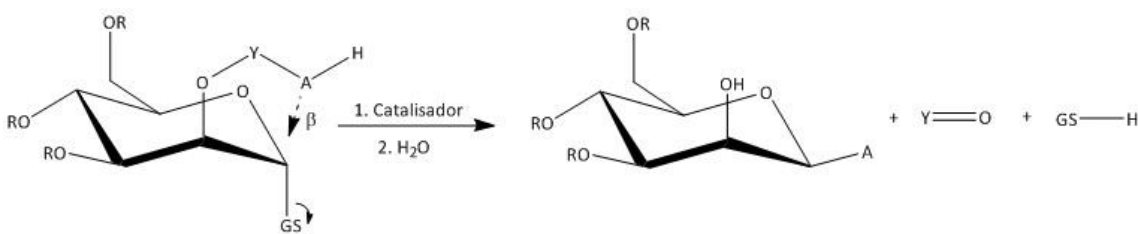

Figura 16 – Glicolisação intramolecular baseada no substituinte funcional (adaptado de Ali, 2003).

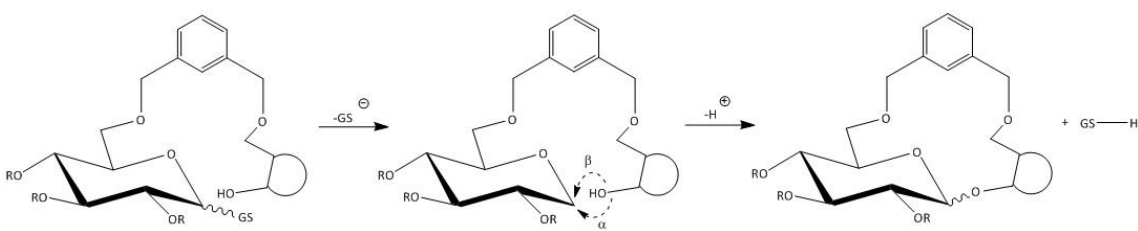

Figura 17 – Mecanismo proposto para a glicolisação intramolecular baseada num espaçador rígido (adaptado de Müller *et al.*, 1999).

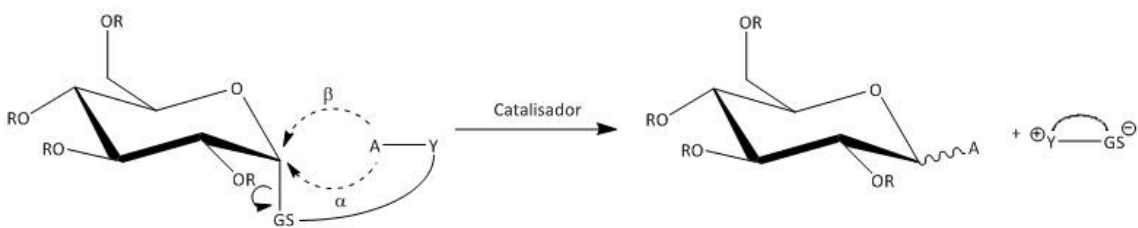

Figura 18 – Glicolisação intramolecular baseada no grupo de saída (adaptado de Scheffler *et al.*, 1997).

Com base na glicolisação por intermédio de um espaçador rígido (Huchel *et al.*, 1998), Schmidt *et al.* (Müller *et al.*, 1999) sintetizaram com sucesso dissacáridos com recurso a resíduos de m-xileno como espaçador, com a estereoseletividade determinada pelo tamanho do anel (de 14 ou 15 elementos) e pela configuração do resíduo aceitador dentro do anel macrocíclico.

### 4.2.3. Método de Fischer-Helferich

O método de Fischer-Helferich baseia-se na substituição direta do oxigénio do carbono anomérico. É um método análogo ao de Koenigs-Knorr, mas sem a utilização de sais de metais pesados, recorrendo apenas a uma catálise ácida de modo a permitir a ligação ao oxigénio do aceitador glicosil.

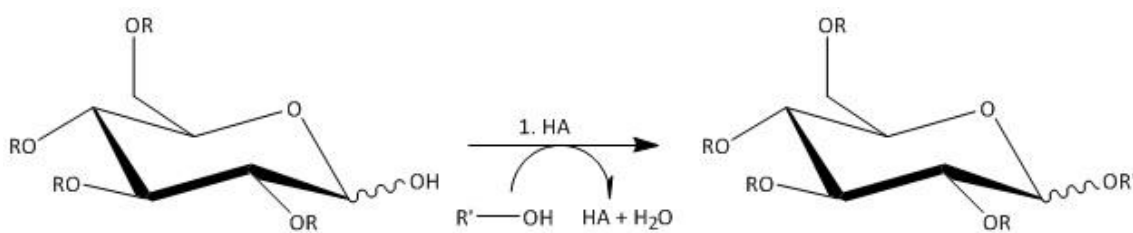

Figura 19 – Glicolisação de Fischer-Helferich, onde R= alquil, acil; R'= alquil, aril, heteroaril; HA= ácidos de Lewis (adaptado de Zhu e Schmidt, 2009).

É um método reversível logo a sua aplicabilidade é limitada e inadequada à síntese de oligosacáridos ou glicósidos complexos. Para uma troca irreversível (assim como no método de Koenigs-Knorr) do oxigénio do carbono anomérico, é necessária uma pré-ativação do centro anomérico pela introdução dum bom grupo de saída.

### 4.2.4. Método de Koenigs-Knorr

Este é o método de glicolisação mais antigo mas é, no entanto, ainda utilizado atualmente. Neste método os dadores glicosilo são, tipicamente, cloretos ou brometos ativados por sais de mercúrio ou prata, ou mais recentemente fluoretos (fluoreto de hidrogénio, dietilaminotrifluoreto de enxofre, sais e metais de fluoreto) (Foster e Westwood, 1973; Yokoyama, 2000).

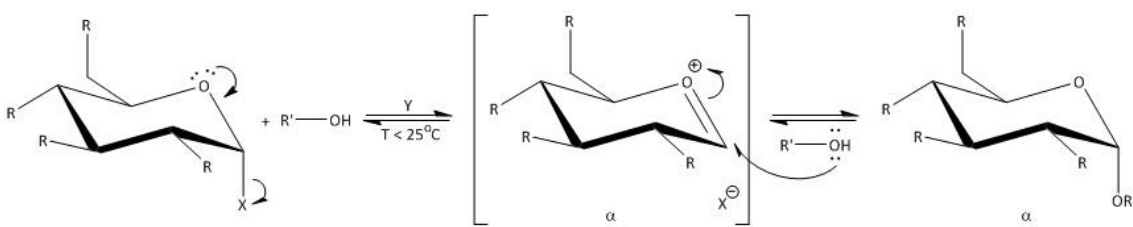

Figura 20 – Glicolisação de Knoenigs-Knorr com recurso a sais, onde R= *O*-alquil, *O*-acil, alquil, acil; R'= alquil, aril, heteroaril; Y= sais de metais pesados, ácidos de Lewis ou catalisadores de transferência de fase (adaptado de Kürti e Czakó, 2005).

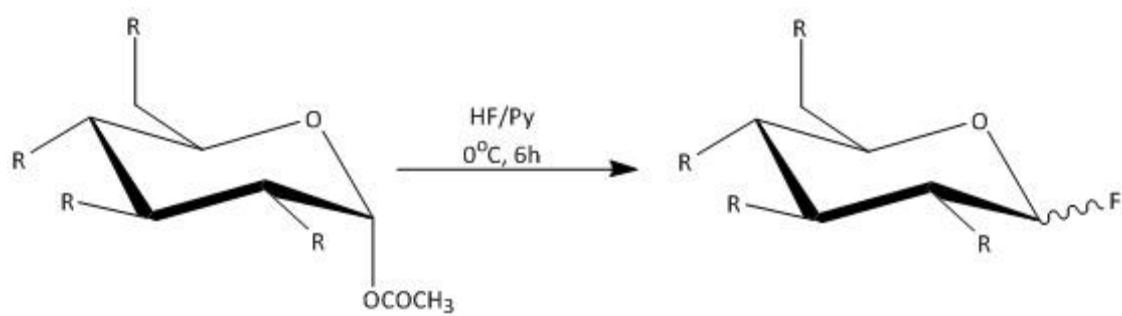

Figura 21 – Glicolisação de Knoenigs-Knorr com recurso a fluoretos, onde R= *O* -alquil, *O* -acil, alquil, acil (adaptado de Yokoyama, 2000).

No entanto, o método de Koenigs-Knorr tem algumas desvantagens: a preparação dos haletos glicosídicos é complexa e estes podem sofrer hidrólise ou eliminação-1,2. Alguns são também termicamente instáveis e são necessárias quantidades estequiométricas de promotor de modo a garantir o sucesso da reação (Kürti e Czakó, 2005).

*4.2.5. Método do Tricloroacetimideto*

Este é um método desenvolvido por R. R. Schmidt e J. Michel, que evita o uso de sais de metais pesados (por oposição ao método descrito no ponto anterior) e permite a obtenção de dadores glicosídicos suficientemente estáveis. São ativados com quantidades catalíticas de ácidos de Lewis, como por exemplo $BF_3 \cdot OEt_2$, ou um catalisador ácido mais forte como o TMSOTf (Ali, 2003).

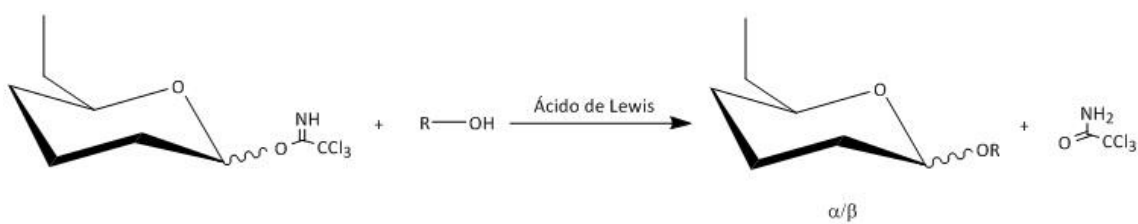

Figura 22 – Glicolisação via tricloroacetimidetos.

O controlo anomérico da ligação *O*-glicosídica depende da configuração anomérica do dador glicosídico, sendo que a configuração $\beta$ é obtida com a utilização de $K_2CO_3$ como base (com controlo cinético, o que permite obter o produto de menor energia de ativação e portanto, mais rapidamente), enquanto a conformação α é obtida com o uso de NaH, $Cs_2CO_3$ ou KOH com um catalisador de transferência de fase (por controlo termodinâmico, o que permite obter um produto mais estável), o que faz deste um método relativamente simples mas que em condições normais produz resultados bastante promissores.

Na síntese dos imidetos, sob catálise básica e por ação do excesso de carga negativa no oxigénio após a saída de um protão, verifica-se a adição direta do grupo tricloroacetimideto.

O elevado poder atraídor de eletrões do grupo triclorometilo facilita a formação do ião oxocarbénio no centro anomérico que, sob catálise ácida (ácido ou ácido de Lewis) leva à saída do grupo tricloroacetimideto. Como este grupo não demonstra propriedades ácidas ou básicas nas condições reacionais, torna possível a sua catálise ácida (com recurso a, por exemplo, $BF_3.OEt_2$), e a expressão das suas boas propriedades dador-glicosídicas, com a formação do aduto *O*-glicósido entre o dador glicosídico.

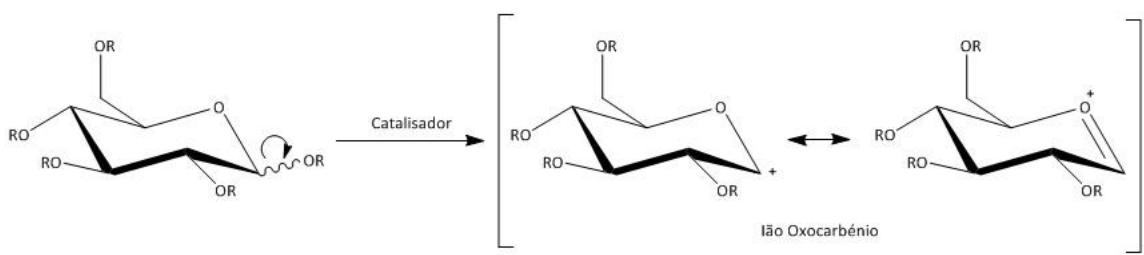

Figura 23 – Possível mecanismo de reação da formação do ião oxocarbénio (adaptado de Banait e Jenks, 1991).

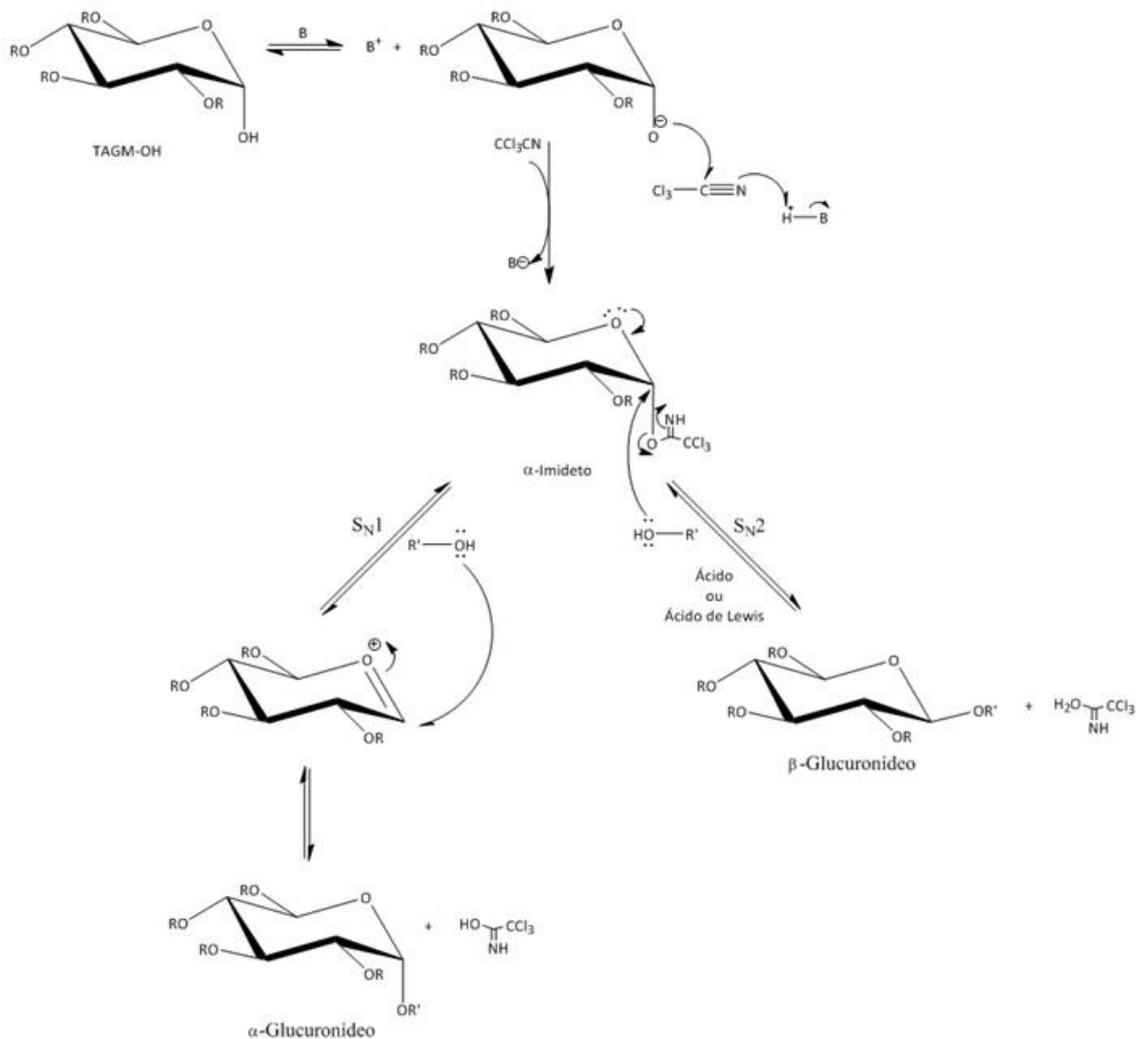

Figura 24 – Mecanismo de síntese proposto e previsto para a síntese do glucoronídeo, onde B= base; R= alquil, acil; R'= alquil, aril, heteroaril (adaptado de Zhu e Schmidt, 2009).

## 5. Conclusão e observações finais

São vastos os trabalhos e ensaios científicos que apontam para uma relação entre a ingestão de azeite – e seus constituintes – e uma menor incidência de diversos tipos de doenças. Nas últimas décadas são notórios diversos indícios que apontam para um benefício regular desta fonte de gordura quando inserida num regime alimentar saudável.

Esses benefícios na saúde não são atribuíveis a um componente específico mas, alguns estudos apontam para que uma parte desses benefícios advenha dos compostos polifenólicos. Estes compostos minoritários do azeite possuem propriedades antioxidantes e da mesma forma poderão conferir proteção antioxidante após ingestão e absorção pelo organismo. No entanto, esta atividade biológica é dependente não só da sua absorção pelos órgãos-alvo bem como da sua metabolização.

O estudo da metabolização e do percurso metabólico é da maior importância pois é através destes que se poderão inferir conclusões acerca da ingestão destes compostos. Assim é necessário proceder à sua síntese, quer por via enzimática quer por via de síntese química. A primeira, apesar de configurar um método rápido e simples para obter os metabolitos, apresenta elevados custos económicos, enquanto a segunda – apesar de execução mais extensa – é consideravelmente menos dispendiosa. Assim, esta é a metodologia mais generalizada e implementada e o que a torna economicamente viável, eficaz e vantajosa.

Seria igualmente interessante complementar os estudos já existentes de biodisponibilidade e ensaios antioxidantes de metabolitos dos polifenóis do azeite. Um dos mais interessantes e promissores seria, por exemplo, proceder à síntese dos glucoronídeos do álcool homovanílico – um composto análogo do neurotransmissor dopamina.

## *6. Bibliografia*